\documentclass[a4paper,conference]{IEEEtran}
\IEEEoverridecommandlockouts
\usepackage{cite}
\usepackage{amsmath,amssymb,amsfonts}
\usepackage{graphicx}
\usepackage{textcomp}
\usepackage{xcolor}
\usepackage{booktabs}
\usepackage{multirow}
\usepackage{stfloats}
\usepackage{microtype}
\usepackage{subcaption}
\usepackage{hyperref}
\hypersetup{hidelinks} 
\usepackage{soul}

\makeatletter
\newcommand{\removelatexerror}{\let\@latex@error\@gobble}
\makeatother
\usepackage{algorithm}
\usepackage{algorithmic}

\def\BibTeX{{\rm B\kern-.05em{\sc i\kern-.025em b}\kern-.08em
    T\kern-.1667em\lower.7ex\hbox{E}\kern-.125emX}}
\begin{document}

\title{HiddenSpeaker: Generate Imperceptible Unlearnable Audios for Speaker Verification System\\
}

\author{
\IEEEauthorblockN{Zhisheng Zhang}
\IEEEauthorblockA{\textit{School of Cyberspace Security} \\
\textit{Beijing University of Posts and Telecommunications}\\
Beijing, China \\
zzs2002@bupt.edu.cn}
\and

\IEEEauthorblockN{Pengyang Huang\IEEEauthorrefmark{2} \thanks{\IEEEauthorrefmark{2} Co-first authors.} }
\IEEEauthorblockA{\textit{School of Cyberspace Security} \\
\textit{Beijing University of Posts and Telecommunications}\\
Beijing, China \\
huangpengyang@bupt.edu.cn}
}

\maketitle

\newcommand{\zhisheng}[1]{\textcolor{red}{#1}\PackageWarning{Zhisheng:}{#1!}}
\newcommand{\pengyang}[1]{\textcolor{blue}{#1}\PackageWarning{Pengyang:}{#1!}}

\begin{abstract}
In recent years, the remarkable advancements in deep neural networks have brought tremendous convenience. However, the training process of a highly effective model necessitates a substantial quantity of samples, which brings huge potential threats, like unauthorized exploitation with privacy leakage. In response, we propose a framework named HiddenSpeaker, embedding imperceptible perturbations within the training speech samples and rendering them unlearnable for deep-learning-based speaker verification systems that employ large-scale speakers for efficient training. The HiddenSpeaker utilizes a simplified error-minimizing method named Single-Level Error-Minimizing (SLEM) to generate specific and effective perturbations. Additionally, a hybrid objective function is employed for human perceptual optimization, ensuring the perturbation is indistinguishable from human listeners. We conduct extensive experiments on multiple state-of-the-art (SOTA) models in the speaker verification domain to evaluate HiddenSpeaker. Our results demonstrate that HiddenSpeaker not only deceives the model with unlearnable samples but also enhances the imperceptibility of the perturbations, showcasing strong transferability across different models.

\end{abstract}

\begin{IEEEkeywords}
deep learning, speaker verification, privacy protection, unlearnable examples
\end{IEEEkeywords}

\section{Introduction}
Due to the rapid breakthroughs in deep learning technology, the field of speaker verification has seen tremendous improvements~\cite{desplanques20_interspeech},~\cite{9747384}. While researchers are excited about this progress, concerns about the privacy of personal audio data have emerged. The reason is that training speaker verification models rely heavily on large-scale voice datasets, which contain abundant voiceprint information. Thus, the exploitation of such personal privacy data without the individual's consent can understandably cause dissatisfaction~\cite{10289030}. Notably, datasets like LibriSpeech~\cite{panayotov2015librispeech}, VCTK~\cite{veaux2017english}, and VoxCeleb~\cite{nagrani2020voxceleb},~\cite{chung18b_interspeech}, which have made significant contributions to the speaker verification field, also involve collecting audio from diverse sources without explicit consent, raising questions about the legality and ethicality of using unauthorized data. Although some countries and regions have clear laws and regulations for the protection of personal data, taking proactive measures to protect one's publicly available audio data is also an effective method, which is precisely the direction our research is striving towards.

\begin{figure}[htbp]
\centerline{\includegraphics[width=0.5\textwidth]{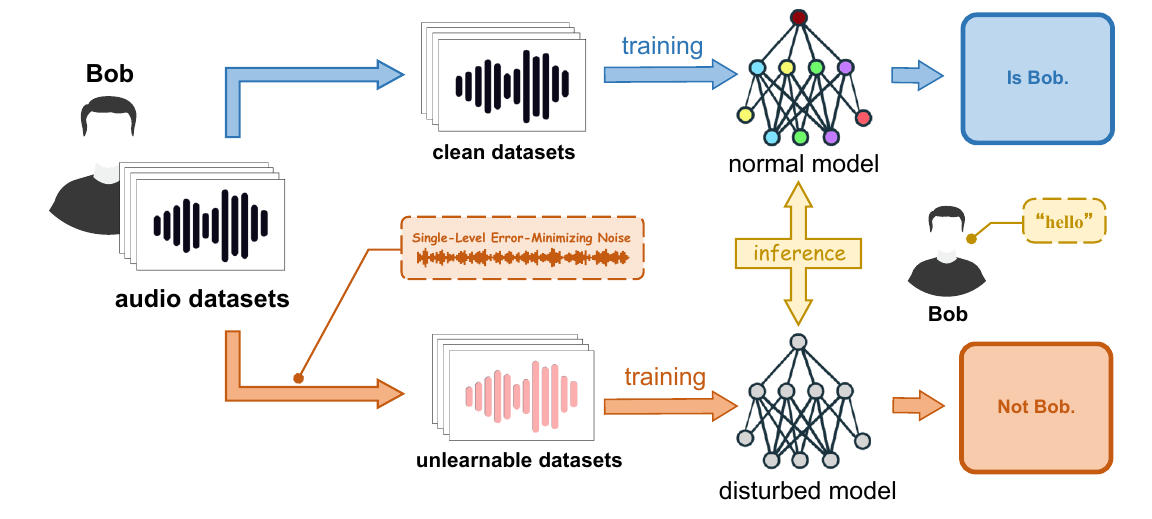}}
\caption{When users upload unprotected audio files to the internet, they become accessible for training purposes. The Single-Level Error-Minimizing method that the HiddenSpeaker system uses, injects noise into raw audio to render these audio datasets unlearnable, thereby disrupting model training effectiveness.}
\label{fig1}
\end{figure}

Efforts to protect privacy in model training data have been ongoing. Tarun et al.~\cite{tarun2023fast} proposed a method for image models that achieves machine unlearning through error-maximizing noise generation and impair-repair weight manipulation. In the audio domain, to discourage hackers from proactively collecting people's voice data, Ge et al.~\cite{ge2022wavefuzz} proposed a clean-label poisoning attack to prevent adversaries from proactively collecting voice data. Although there are many current works on privacy protection, the protection of training data for speaker verification models is still not perfect, and the results in this field are not ideal. Therefore, we propose a privacy protection system for the speaker verification model named HiddenSpeaker. Building upon the limitation of $\ell_p$ norm that introducing imperceptible perturbations, difficult for a human to distinguish, renders the model unlearnable, we extend this concept from the field of images~\cite{huang2021unlearnable} to the speaker verification domain. HiddenSpeaker is designed to generate protected audio data that is indistinguishable from the original, offering a privacy safeguard for audio information so that it cannot be used for training by speaker verification models. Compared to image data where perturbations are primarily spatial: (1) Audio perturbations must consider both the time and frequency domains, presenting unique challenges in preserving the perceptual quality; (2) Human auditory perception is highly sensitive to even subtle changes in sound than the eye, making the task of adding noise to audio more challenging compared to images. This requires a careful balance in the design of HiddenSpeaker, ensuring that the added noise is effective in preventing model training while remaining undetectable to listeners. The intricacies of audio data, such as its temporal structure and spectral characteristics, demand a nuanced approach to noise addition, one that respects the natural dynamics and rhythms of audio.

Integrating the above challenges, our HiddenSpeaker system consists of two key components. Firstly, we employ a simplified error-minimizing method to generate imperceptible perturbations called Single-Level Error-Minimizing (SLEM), which renders the data unlearnable when embedding to the raw audio as \autoref{fig1}. The SLEM method only utilizes the internal loop, thereby reducing the loss function faster. Then HiddenSpeaker utilizes Perceptual Hybrid Losses (PHL) for noise perceptual optimization, making the added noise hard to detect. PHL incorporates the Short-Time Fourier Transform (STFT)~\cite{benesty2011speech} loss and the Short-Time Objective Intelligibility (STOI)~\cite{jensen2016algorithm} loss. It optimizes the noise in both frequency and time domains, aligning it with human perception. The STFT loss maintains the integrity of the audio's spectral characteristics, while the STOI loss preserves speech intelligibility~\cite{liu2023inplace}.

To make the speaker verification models better able to learn the generated perturbation, we choose to embed noise in the high value of the amplitude range to compensate for the potential decrease in noise interference ability due to noise optimization. In the image domain, it has been shown that adding noise at high frequency has more interference effect than at lower one~\cite{wang2020high}, so in the audio domain, we guess that embedding noise to high values of amplitude would have a similar effect. The experimental results in Section IV subsection C demonstrate the significant effectiveness of this strategy. As the HiddenSpeaker approach balances the impact of noise on the model and the imperceptibility of the noise, the addition of noise does not affect users’ normal usage of audio. We have also verified that the HiddenSpeaker system has a certain level of transferability, meaning the protected audio generated by this approach can work on a certain range of different models.

In summary, our work makes three main contributions:

\begin{itemize} 
\item We propose a privacy protection system named HiddenSpeaker to prevent unauthorized exploitation of voice samples in the speaker verification domain, utilizing the SLEM noise generation scheme. 
\item We introduce the PHL function that fuses STFT and STOI losses, ensuring the resultant noise remains imperceptible. 
\item We verify the effectiveness and transferability of protected audio generated by the HiddenSpeaker system, proving its potential to secure against the SOTA model.
\end{itemize}

\begin{figure*}[htbp]
\centerline{\includegraphics[width=0.9\textwidth]{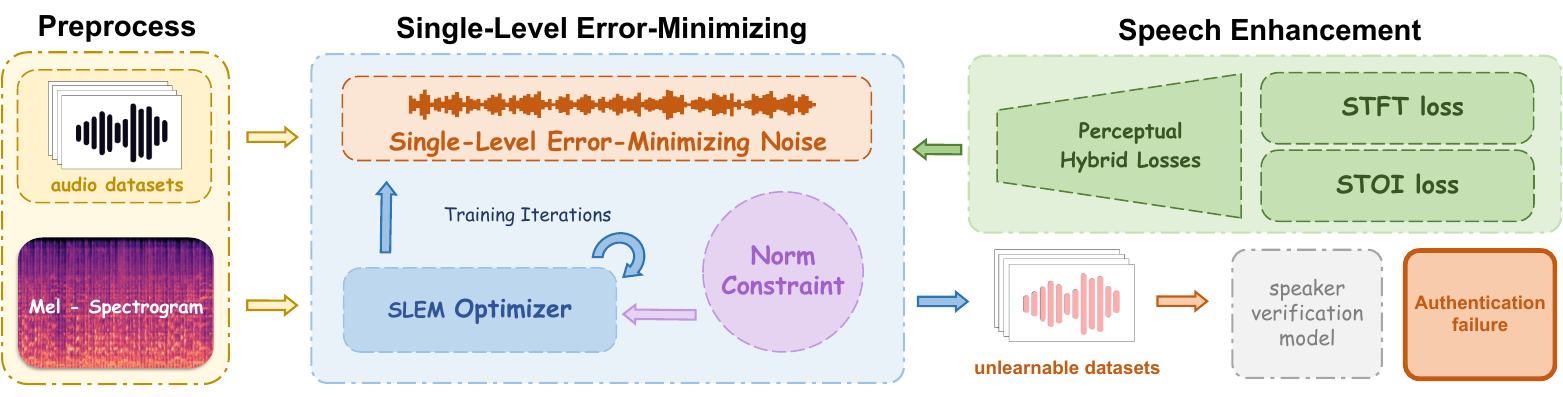}}
\caption{The HiddenSpeaker system workflow operates in two phases. SLEM noise is embedded into the audio in need of protection. Subsequently, a PHL function optimizes this noise, factoring in both STFT and STOI considerations, to maintain auditory indiscernibility.}
\label{fig2}
\end{figure*}

\section{Related Work} In this section, we provide and compare a brief overview of the domains relevant to our study, including privacy protection, poisoning attacks, and speaker recognition.

\subsection{Privacy Protection} The widespread application of deep learning across various domains has heightened concerns for data privacy, as the training of models requires extensive data containing personal information. To protect sensitive information within data, a series of approaches such as differential privacy~\cite{abadi2016deep} and homomorphic encryption~\cite{gilad2016cryptonets} have been proposed. Techniques like homomorphic encryption have also been widely used in the audio domain. For example, Lai et al.~\cite{lai2022efficient} discussed the use of encrypted audio fragile watermarking using homomorphic encryption and SIMD batching in cloud computing. Ezgi Zorarpacı et al.~\cite{zorarpaci2020hybrid} proposed a method that achieves both security and privacy by processing the encrypted training audios and minimizing the privacy leakage of individuals. However, these solutions significantly increase the computational overhead and require customization for different models, and they are not universal. Certainly, more simple and universal solutions exist; techniques such as obfuscation~\cite{wang2021infoscrub} are used to protect media information. Techniques specifically tailored for audio include advanced anonymization of voice characteristics ~\cite{10095173}, and selective noise addition to mask-sensitive audio elements~\cite{mivule2013utilizing}. However, these solutions impair the usability of the audio, making it highly unlikely for users to upload such protected audio to social platforms, which is a problem we hope to solve. Therefore, we are inspired by the image domain's error-minimizing noise schemes~\cite{huang2021unlearnable} aimed at addressing the unauthorized use of personal data in training speaker verification models. The image domain's error-minimizing noise approach has also been transferred to the privacy protection of text data, where Li et al.~\cite{li2023make} proposed a method for extracting simple patterns from unlearnable text and successfully verified the method's usability under multiple models. It is evident that data privacy protection is needed across various domains, and the protection of privacy in training data for speaker verification models is particularly urgent.

\subsection{Poisoning Attack}

Since the concept of poisoning attack was put forward~\cite{barreno2006can}, techniques for poisoning deep learning models have been continually practiced~\cite{biggio2012poisoning},~\cite{carlini2017towards}. Currently, poisoning attacks can be roughly divided into two types~\cite{goldblum2022dataset}, data poisoning and backdoor attacks. Data poisoning~\cite{shafahi2018poison} involves tampering with a certain amount of data to affect the overall accuracy of model inferences. In contrast, backdoor attacks~\cite{guo2023masterkey} maintain normal inference accuracy on benign samples, while reasoning errors in the face of samples with triggers, and the results are often controlled by the attackers. Both types of attacks can affect the availability and stability of the model by adding perturbations to the sample. The same kind of attack works in the audio domain~\cite{10082866}. Aghakhani et al.~\cite{aghakhani2023venomave} presented the first data poisoning attack in the audio domain, named VENOMAVE, which demonstrated the practical feasibility of this attack and underscores the need to consider data poisoning attacks as a real threat to Automatic Speech Recognition systems. Ge et al.~\cite{ge2022wavefuzz} explored a novel clean-label poisoning attack called WaveFuzz, which operated by perturbing audio data to generate poisoned frequency features and significantly degrade the performance of audio intelligence systems, including speaker recognition and speech command recognition systems. Our work has similarities with poisoning attacks but with a contrary intention. We aim to render the model difficult to learn useful information from data by embedding perturbations into training samples, which are imperceptible to humans, to decrease the training effect.

\subsection{Speaker Recognition}

Studies on speaker recognition technology have a long history, among which different technical schemes can be divided into two types~\cite{chung18b_interspeech}, traditional methods and deep learning methods. Among traditional methods, some solutions have been able to solve the problems of channel or session variability~\cite{reynolds2000speaker}. One standout approach in traditional methods is the i-vector~\cite{alam2011multi} scheme, which uses a low-dimensional speaker and channel-dependent space defined by factor analysis. Subsequent research has further refined this approach by introducing enhancements such as Probabilistic Linear Discriminant Analysis (PLDA) and a heavy-tailed variant of PLDA for scoring i-vectors~\cite{matvejka2011full}. 

However, traditional methods still have some drawbacks, particularly in handling complex scenarios where their performance may be unsatisfactory. In contrast, deep learning methods can cope with the change and complexity of the scene, and the application of representation learning~\cite{chen2022large} and end-to-end learning~\cite{variani2014deep} schemes makes speaker recognition technology more stable and more mature.

Studies on speaker recognition can be subdivided into several different tasks such as speaker verification~\cite{bai2021speaker}, speaker identification, and diarization, among which speaker verification is more relevant to our study. For example, ECAPA-TDNN~\cite{desplanques20_interspeech}, which has reached the SOTA performance, has made contributions to the development of this field, and this model is also one of our experimental models.

\section{Method}
To reduce the loss function by embedding noise more quickly, we simplify the initial bi-level error-minimizing method~\cite{huang2021unlearnable} and adopt the inner loop, which can be termed the Single-Level Error-Minimizing (SLEM) method. Building on this, to address the issue of SLEM noise being detectable, we employ a Perceptual Hybrid Losses function for audio enhancement to minimize the impact of noise addition on the original audio, thereby improving the quality and auditory experience of the generated unlearnable audio.


\textbf{Assumption and capability.} We assume that the model trainer can collect publicly available voice information from the internet and utilize these data for training a speaker verification model without authorization. The objective of the audio protector is to prevent unauthorized entities from actively collecting users' voice data for speaker verification model training by generating protected data. Audio protectors possess the capability to add perturbations into audio with the HiddenSpeaker system and then upload the protected audio, which sounds normal compared to unprotected audio, on social platforms. The workflow of the system is shown as \autoref{fig2}. At the same time, we assume that the audio protector is knowledgeable about the model's information, and the model primarily belongs to the ECAPA-TDNN or its related models.

In this section, we provide a detailed introduction to the SLEM method and the PHL function, which offers a clearer understanding of how SLEM achieves perturbation to influence model training, and we also explain how the PHL function enhances the concealment of the perturbation.

\subsection{Single-Level Error-Minimizing}\label{AA}

In research on the computer vision domain, a type of data protection by embedding error-minimizing noises was proposed on deep neural networks~\cite{huang2021unlearnable}. They crafted the error-minimizing noises by a bi-level optimization that the inner loop is for optimizing the noise, while the outer loop is for parameters $\theta$ of the targeted model $f$. The bi-level error-minimizing optimization problem can be described as:

\begin{equation}
\min_\theta \frac{1}{N}\sum\limits_{i=1}^N \min_{\delta_i}[\mathcal{L}(f_\theta(x_i+\delta_i), y_i)], \label{eq1}
\end{equation}
where $N$ represents the number of the data to be protected, $\delta_i$ is the i-th perturbation which is bounded by $\ell_p$ norm as $||\delta_i||_p\le \epsilon$ with controlling $\epsilon$ for generating slight one. $(x_i, y_i)$ denotes input data and its label.

The formula \ref{eq1} brings much computational complexity and time when optimizing both a noise generator and model parameters. Meanwhile, it is more difficult to simultaneously optimize the time delay neural network $g(\cdot )$ on the speaker verification domain to reach the set loss function value with the bi-level optimizer for noise generation. Based on the above, we utilize a Single-Level Error-Minimizing noise generator which eliminates the optimization of parameters of the model and preservers the internal loop, thereby faster reducing the loss function by adding noise. Based on this, the single-level error-minimizing problem can be expressed by: 

\begin{equation}
\min_{\delta_{i}} \frac{1}{N}\sum\limits_{i=1}^{N}\mathcal{L}(g(x_i+\delta_i), y_i). \label{eq2}
\end{equation}

Using the SLEM method, we generated ``sample-wise'' noise for different speech samples and ``speaker-wise'' noise based on the voiceprints of different speakers, both of which can provide fine voiceprint protection for audios in the current SOTA speaker verification model in Section IV.

\subsection{Speech Enhancement}\label{AA}
To improve the noise quality using SLEM, we utilize the PHL function for optimizing the perception of the noises in frequency and time domains. We exploit the distance between the original and protected examples after STFT as loss function and the standard STOI loss function for improving the comprehension for humans which will be introduced in the following part. The total losses can be described as:

\begin{equation}
\mathcal{L}_{total} = \alpha \mathcal{L}_{Arc} +\beta \mathcal{L}_{stft} + \gamma \mathcal{L}_{stoi},\label{eq3}
\end{equation}

where $L_{Arc}$, $L_{stft}$ and $L_{stoi}$ represent the loss Additive Angular Margin (AAM)~\cite{deng2019arcface} loss function for speaker verification model, STFT loss function and STOI loss function for noise optimizing. $\alpha, \beta, \gamma$ are weight coefficients. The STFT loss and STOI loss are described in detail below.


The SOTA models in the speaker verification domain usually take the first step to extract the time domain and frequency domain features by STFT. On this basis, to make the added noise more consistent with the spectral characteristics of the audio signal and maintain the alignment of the time-frequency information of the audio, we use the STFT loss function for describing the difference between the original and unlearnable audio after STFT. The function can be formulated as:

\begin{equation}
\mathcal{L}_{stft} = \left|\text{STFT}(x_i + \delta_i) -\text{STFT}(x_i)\right|_2\label{eq4},
\end{equation}
where $\text{STFT}(\cdot)$ is the Short-Time Fourier Transform for input audio data. $|\cdot|_2$ represents the $\ell _2$ distance.

STOI score is a measure that can predict the clarity of noise or processed speech, which takes into account human auditory perception and ranges from 0 to 1. A higher STOI value indicates better speech clarity. Therefore, we exploit it as our perceptual loss. We need to calculate the STOI score first, which is achieved by grouping DFC bins after removing silent frames and performing STFT. A total of 15 one-third octave bands are used, with the lowest center frequency set to 150Hz, and the center frequency of the higher one-third octave band set to 4.3kHz. The short-time envelope vector of clean speech is shown as:
\begin{equation}
x_{j,m} = [X_j (m-N+1), X_j(m-N+2), ..., X_j(m)]^T,\label{eq5}
\end{equation}
where $m$ is the index of the frame and $j\in \{1, 2, ..., 15 \}$ is the index of the third-harmonic band. $X$ is the obtained third-harmonic band.

Subsequently, the normalized and limited time envelope of the degraded speech is represented as $x_{j,m}$. The STOI calculates the average of the intermediate intelligibility across all frequency bands and towns, while the intermediate intelligibility metric is defined as the correlation coefficient between two-time envelopes. The STOI score is formulated as:

\begin{equation}
f_{stoi}(X_m, X'_m) = \sum\limits_{j=1}^J\frac{(x_{j,m} - \mu_{x_{j,m}}) (x'_{j,m} - \mu_{x'_{j,m}})}{||x_{j,m} - \mu_{x_{j,m}}||_2||x'_{j,m} - \mu_{x'_{j,m}}||_2},\label{eq6}
\end{equation}
where $\mu(\cdot )$ represents the sample mean of the input vectors and $J$ is the total number of the one-third octave bands. Based on this, by setting an appropriate weight parameter $\lambda$, the STOI loss function is defined as:

\begin{equation}
\mathcal{L}_{stoi} = (1-f_{stoi}(X_m, X'_m))^2 + \lambda \left(||X^J_m-X'^J_m||_1 / J\right).\label{eq7}
\end{equation}

To compensate for the potential attenuation of SLEM capabilities due to noise optimization, we attempt to incorporate the noise into the high value of the amplitude components of the audio. Since a well-studied feature of models is their tendency to capture the higher value of the amplitude components in the data ~\cite{wang2020high}, we infer that embedding noise to the high value of the amplitude might have a more disruptive effect than adding it to the low value of the amplitude. Therefore, to achieve a better unlearnable effect, we choose to introduce noise in the high value of the amplitude. The experiments demonstrated that this strategy yielded effective results.

\section{Experiments}
In this section, we design experiments to demonstrate that audio data processed by HiddenSpeaker has remarkable performance when training on ECAPA-TDNN or its related models. First, we list the details of the experimental design scheme, and then We describe four experiments to verify the effectiveness of the SLEM method, extreme conditions, the PHL method, and noise transferability.

\subsection{Experimental Setup}\label{AA}
We outline the experimental settings used in our experiments.

\textbf{Datasets.} We train models on the complete VoxCeleb1~\cite{nagrani2020voxceleb} train dataset and the first 2,000 speakers dataset of VoxCeleb2~\cite{chung18b_interspeech} about 340,827 utterances, both of which are recognized in the field of speaker verification. VoxCeleb1 was curated using the pipeline to assemble a collection of hundreds of thousands of real-world utterances from over 1,000 celebrities. For evaluation, we select the VoxCeleb1-O test dataset. It ensures a standardized assessment of model performance, consisting of selected audio samples that present challenges for speaker verification systems.

\textbf{Experimental Details.} Our experiments are all conducted on a NIVIDA Tesla V100 GPU with 32 GB memory and the hyperparameter $\alpha$ is set to 1, and $\beta$ and $\gamma$ are set to 0.005 and 0.01, respectively. In the noise process, we optimize the perturbation at a fixed patch and train the model as the same set. This decision is made to maintain methodological consistency. We directly add noise to the original audio and set $\epsilon$ as 0.005 to obtain a better sound with the effectiveness of unlearnability. After obtaining the unlearnable data, we train the model for 30 epochs to fully learn the features. 

\textbf{Models.} The foundational model selected for our experiments is the ECAPA-TDNN~\cite{desplanques20_interspeech}, a prevailing voiceprint recognition model, achieving first place in the VoxCeleb Speaker Recognition Challenge in the same year of its proposal. The ECAPA-TDNNs model has demonstrated SOTA performance in the field of speaker verification. In our experiments, we also employ the models DS-TDNN~\cite{li2023dual}, MFA-Comformer~\cite{zhang22h_interspeech}, and MFA-TDNN~\cite{liu2022mfa}, all of which also achieve SOTA performance.

\textbf{Metric.} We quantify our evaluation using Equal Error Rate (EER) for unlearnability, where lower EER values indicate higher accuracy. We also employ the Detection Cost Function (DCF) to calculate minimum DCF (minDCF) and a smaller value of minDCF indicates a better performance of the model. For evaluating perturbation imperceptibility, we employ Mean Squared Error (MSE)($e^{-6}$), and Signal-to-Noise Ratio (SNR). MSE quantifies signal variance, where lower values indicate minimal impact. Higher SNR denotes clearer signals, while lower SNR may indicate increased noise introduction.

\begin{table*}[t]
    \centering
    \caption{Comparison of ECAPA-TDNN and its related models trained on clean, random noises added, and SLEM noises added dataset and the evaluation metrics of training are EER and MinDCF.}
    \resizebox{0.95\textwidth}{!}{
        \begin{tabular}{ccccccccccc}
            \toprule
            \multirow{2}[1]{*}{Dataset} 
            & \multirow{2}[1]{*}{Method} 
            & \multicolumn{2}{c}{ECAPA-TDNN}
            & \multicolumn{2}{c}{DS-TDNN}
            & \multicolumn{2}{c}{MFA-Conformer}
            & \multicolumn{2}{c}{MFA-TDNN}\\
            \cmidrule(lr){3-4}\cmidrule(lr){5-6}\cmidrule(lr){7-8}\cmidrule(lr){9-10}
            & & EER(\%)($\downarrow$) & MinDCF($\downarrow$) 
            & EER(\%)($\downarrow$) & MinDCF($\downarrow$) 
            & EER(\%)($\downarrow$) & MinDCF($\downarrow$) 
            & EER(\%)($\downarrow$) & MinDCF($\downarrow$) \\
            \midrule
            \multirow{3}{*}{$D_1$}
             & clean & 4.830 & 0.301 & 5.250 & 0.352
                   & 5.292 & 0.347 & 4.936 & 0.329\\
            & random noise & 5.255 & 0.342 & 5.986 & 0.363
                        & 5.944 & 0.385 & 5.760 & 0.362\\
            & SLEM &\textbf{24.919}& \textbf{0.947} & \textbf{22.402} &\textbf{0.908}
                & \textbf{25.105} & \textbf{0.913} & \textbf{28.545}& \textbf{0.908}\\

            \midrule
            \multirow{3}{*}{$D_2$}
            & clean & 3.674 & 0.249 & 3.759 & 0.223 
                   & 4.095 & 0.251 & 3.808 & 0.241\\
            & random noise & 3.867 & 0.272 & 4.648 & 0.297
                   &4.822  &0.304  & 4.393 &  0.274\\
            & SLEM  & \textbf{23.595} & \textbf{0.921} & \textbf{21.310} &\textbf{0.900}
                  &\textbf{26.013} &\textbf{0.915} & \textbf{26.615} & \textbf{0.962}\\
            
            \bottomrule
            \multicolumn{10}{l}{${D_1}$ and ${D_2}$ represent the dataset VoxCeleb1 and VoxCeleb2, respectively.}
        \end{tabular}
        \label{tab:1}
    }
\end{table*}

\subsection{Effectiveness of Single-Level Error-Minimizing Noise}\label{AA}
To assess the effectiveness of SLEM noise, we conduct comparative experiments on models including ECAPA-TDNN, DS-TDNN, MFA-Conformer, and MFA-TDNN as introduced earlier. We employ different models to generate noise and train to verify the usability of the method.

\autoref{tab:1} presents the experimental results, detailing the training performance of different models on both clean and SLEM-protected datasets. The SLEM method consistently generates unlearnable data across different datasets, effectively degrading the training performance of the models. In the experiment, the noise optimization time performed by the SLEM method is also within the acceptable range. Take the noise optimization performed on the VoxCeleb1 dataset which contains 148,641 samples for the ECAPA-TDNN model as an example, noise optimization on the entire dataset with our device takes about 183 minutes, and optimizing individual samples takes about 70 milliseconds on average. It is worth noting that the recorded experimental results for EER and minDCF are the lowest values observed during the training period. Even so, such values are still significantly high,  all models trained on SLEM-protected datasets exhibit a significant increase in both EER and minDCF compared to training on clean datasets. The increase in EER values, exceeding 400\% on average, highlights the disruptive impact of SLEM on model training. For example, under the VoxCeleb1 training set, the MFA-TDNN model experiences a remarkable 478\% increase in EER, reaching 28.545\%. The minDCF values further support this observation, rising from around 0.3 during training on clean datasets to above 0.9 for models trained with datasets via SLEM. When the model's EER reaches such high values, it can be considered that the model's training is ineffective and the SLEM method is successful in protection. Take ECAPA-TDNN as an example, the embeddings calculated by ECAPA-TDNN at this time are not accurate with a failed speaker verification.

These results affirm that SLEM effectively disrupts model training across datasets and models, validating its efficacy. The observed disruptive impact on EER and minDCF values provides a strong foundation for subsequent experiments.

\subsection{Noises Analyses}\label{AA}
Cause that in reality world it is impractical to obtain a large number of user's speeches for perturbations generation. So in a further exploration of the capabilities of SLEM, we experiment with more methods of perturbation generation and assess the effectiveness of SLEM noise under extreme conditions.

Therefore, we intentionally select one sample from each of the 1,211 speakers in the VoxCeleb1 dataset to create a subset referred to as ``speaker-wise''. In this setup, the ``speaker-wise'' dataset is used to generate perturbation and subsequently transfer it to the entire VoxCeleb1 dataset with the corresponding speaker, which is used for the training of the ECAPA-TDNN model.

\begin{figure*}[!t]
\centering
\subfloat[EER values over epochs]{
		\includegraphics[scale=0.5]{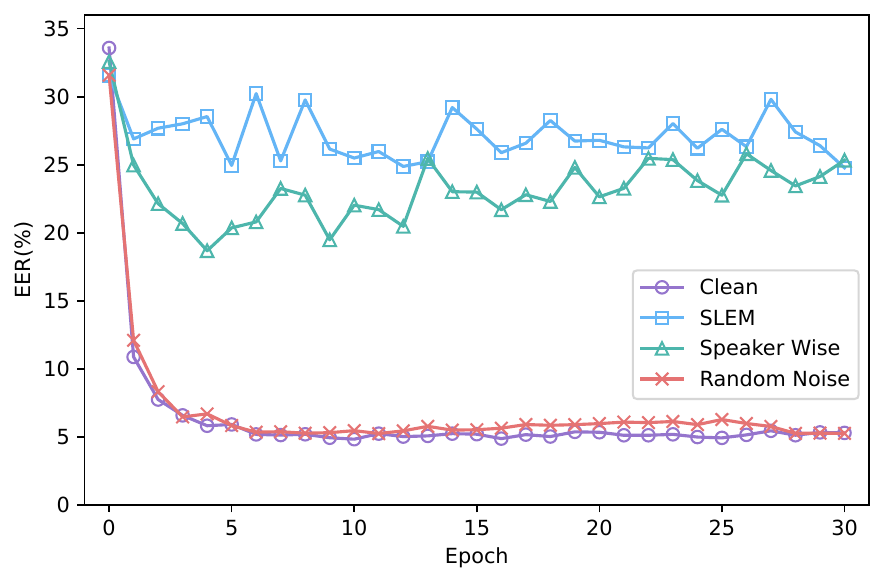}}
\subfloat[minDCF values over epochs]{
		\includegraphics[scale=0.5]{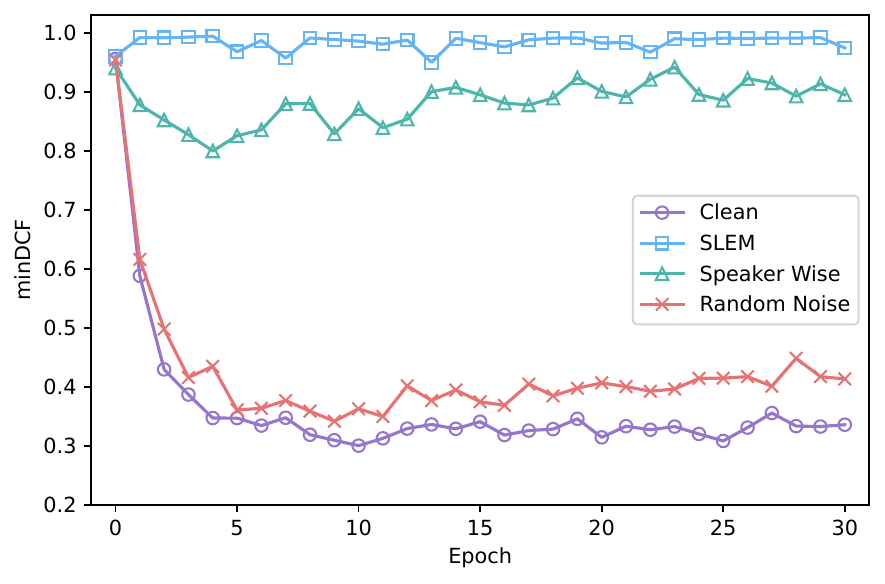}}
\caption{EER and minDCF values over epochs when different types of noises are added to the complete VoxCeleb1 dataset for ECAPA-TDNN model training, and clean VoxCeleb1 samples without added noise as a control group.}
\label{fig3}
\end{figure*}

\autoref{fig3} illustrates the evolution of EER and minDCF values over epochs as different noise generation methods are applied to the complete VoxCeleb1 dataset for ECAPA-TDNN model training. The experiment compares samples of SLEM-protected, ``speaker-wise'' generated, random Gaussian noise added, and unprotected clean datasets, respectively. By comparing the training effects of each sample on the ECAPA-TDNN model, we assess the effectiveness of the noise generated under speaker-wise conditions. Surprisingly, the reduction in EER achieved with speaker-wise SLEM, an extreme condition, is only approximately 1.2\% lower than the EER obtained with the normal SLEM method using the complete dataset, reaching an EER of 20.014\% after 30 epochs, with a particularly alarming EER of 25.332\% at the 30th epoch, indicating the speaker-wise SLEM experiment significantly interfered with training effectiveness. Furthermore, a minDCF of approximately 0.895 supports the efficacy of this extreme approach. The speaker-wise method closely follows the normal SLEM curve and becomes more similar as epochs progress, indicating that the noise generated under speaker-wise conditions is very similar in effect to the noise generated under normal conditions by the SLEM method, making the verification of the speaker fail. Additionally, attempts to train the model using random noise added to the dataset showed minimal interference with training, closer to the performance on clean datasets, indicating that not all noise interferes with model training, making the experiment more rigorous.

In conclusion, the speaker-wise SLEM experiment reveals that SLEM exceeds initial expectations, inducing an unlearnable effect even with noise generated from a small subset. This suggests SLEM's adaptability to challenging situations.

\begin{table}[t]
    \centering
    \caption{Comparison of audio sample quality generated using SLEM method and PSLEM method for four models.}
    \resizebox{0.4\textwidth}{!}{
        
        \begin{tabular}{cccccc}
            \toprule
            \multirow{3}{*}{Model} & \multirow{3}{*}{\# Params} & \multicolumn{2}{c}{SLEM}  \\
            \cmidrule(lr){3-4}
            & &SNR($\uparrow$) & MSE($\downarrow$)  \\
            \midrule
             ECAPA-TDNN &14.73 M &22.190	& 9.498  \\
            DS-TDNN &20.97 M &22.515 & 8.692   \\
            MFA-Conformer &17.87 M &22.137 & 9.591  \\
            MFA-TDNN &20.84 M &22.397 & 8.988   \\
            \midrule
            \multirow{3}{*}{Model} & \multirow{3}{*}{\# Params}& \multicolumn{2}{c}{PSLEM}  \\
            \cmidrule(lr){3-4}
            & &SNR($\uparrow$) & MSE($\downarrow$)  \\
            \midrule
            ECAPA-TDNN & 14.73 M&\textbf{26.589} & \textbf{3.494}   \\
            DS-TDNN &20.97 M &\textbf{26.457} & \textbf{3.596}   \\
            MFA-Conformer & 17.87 M&\textbf{26.518} & \textbf{3.540}   \\
            MFA-TDNN &20.84 M &\textbf{26.583} & \textbf{3.511}  \\
            \bottomrule
        \end{tabular}
        \label{tab:2}
    }
\end{table}

\subsection{Effectiveness of Perceptual Hybrid Losses Function}\label{AA}
To assess the impact of the PHL function on enhancing the imperceptibility of SLEM noise, we conduct a series of experiments comparing the audio sample quality generated using the standard SLEM method with SLEM optimized through the PHL (PSLEM) for four different models. We evaluate the audio quality using SNR, and MSE, providing a comprehensive analysis of PHL function's efficacy.

The VoxCeleb1 dataset serves as the training set. The experimental results are presented in \autoref{tab:2}. Notably, without the PHL optimization, the SNR values for the generated noisy datasets by all four models are approximately 22. However, with the application of PHL optimization, an improvement can be observed, with a general increase of around 20\% in SNR values for the HiddenSpeaker-protected datasets generated by each model. Furthermore, the respective MSE values were reduced by approximately 60\%, indicating a substantial reduction in noise distortion due to the PHL optimization. These figures suggest that the noise is nearly imperceptible to ears. Importantly, the interference effects on model training are nearly consistent between PSLEM and SLEM. For example, using the SLEM method, the EER for the MFA-TDNN model is 28.545\% while using the PSLEM method, it slightly decreases to 26.131\%, with very similar results that still prevent the trained MFA-TDNN model from correctly completing the speaker verification task. The PHL optimization does not significantly alter the capability of SLEM to interfere with model training. This result validates our strategy of embedding noise to the high value of the amplitude as effective.

\begin{figure*}[htbp]
\centerline{\includegraphics[width=1.0\textwidth]{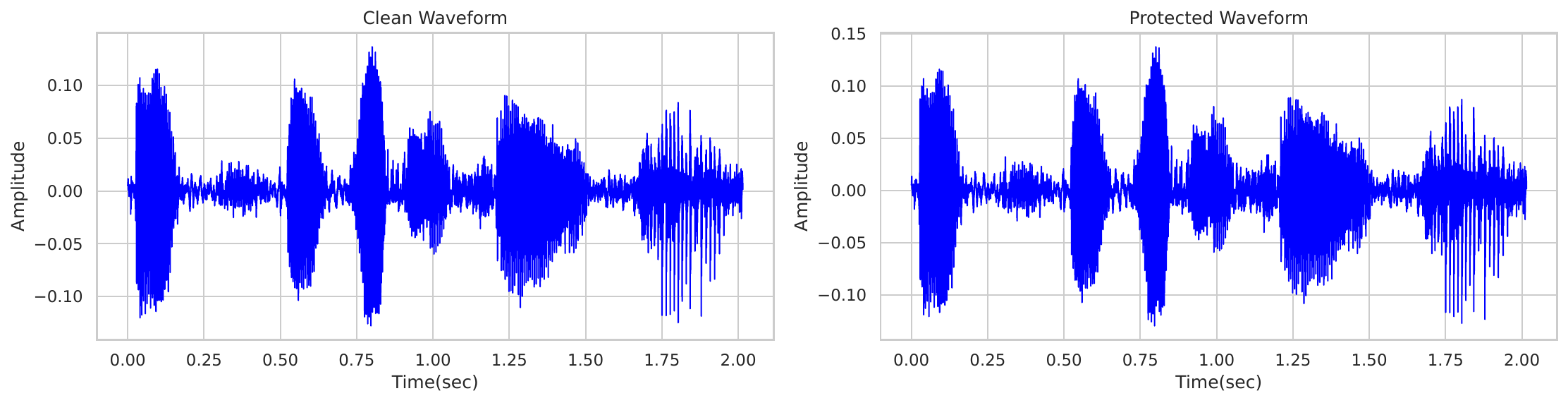}}
\caption{The visual comparison between the original audio waveform and the HiddenSpeaker-protected waveform. It can be observed that visually, there is no clear difference between the two waveforms.}
\label{fig4}
\end{figure*}

To visually demonstrate the effectiveness of PHL optimization, we perform visualizations on the audio data. \autoref{fig4} displays the waveform comparison between clean audio data and the protected sample by HiddenSpeaker. The two waveforms are challenging to differentiate, indicating that the PSLEM-protected speech data can successfully convey information without exhibiting noticeable differences.

In summary, the results affirm that the PSLEM noise elevates the quality of the HiddenSpeaker-protected dataset, making it closely resemble a noise-free state. This further validates the PHL as an optimization technique, substantially aiding in enhancing the imperceptibility of noise for privacy protection in the audio domain.

\begin{table}[H]
    \centering
    \caption{ECAPA-TDNN, DS-TDNN, and MFA-Conformer trained on the protected dataset generated by MFA-TDNN.}
        \begin{tabular}{ccccccccc}
            \toprule
             Model & EER(\%)($\downarrow$) & MinDCF($\downarrow$) \\
            \midrule
             ECAPA-TDNN & 26.381 & 0.966  \\
            DS-TDNN & 23.457 & 0.901 \\
            MFA-Conformer & 23.717 & 0.913\\
            
            \bottomrule
        \end{tabular}
        \label{tab:3}
\end{table}

\subsection{Transferability Analyses}\label{AA}
We consider a crucial question: \textit{Can the SLEM noise generated by a surrogate model be effectively transferred to disrupt the training of other models? }

In reality, it is often challenging to know in advance which model the trainer will employ for training after obtaining user data, so we cannot guarantee that we can choose the corresponding model for noise generation based on the type of model trained. Therefore, the transferability of the noise becomes crucial. To explore this hypothesis, we experiment to see if we could use one model to generate SLEM noise, apply this noise to the complete VoxCeleb1 dataset, and then employ the dataset to train different models checking if it can also perturb the training of these models.

As seen in \autoref{tab:1}, the noise generated by MFA-TDNN demonstrates the most effective interference with model training. Consequently, we select MFA-TDNN as the surrogate model, while the other three models directly employ the HiddenSpeaker-protected dataset generated by MFA-TDNN for training. \autoref{tab:3} presents the results of this experiment. Remarkably, the three models, when trained on the HiddenSpeaker-protected dataset generated by MFA-TDNN, exhibit significant interference effects. The EER for all three models exceeds 23\% after 30 epochs, with ECAPA-TDNN reaching a best EER of 26.381\%. Such a value means that the trained models cannot perform speaker verification tasks correctly. The unlearnable effect on the DS-TDNN model using the HiddenSpeaker-protected dataset from MFA-TDNN surpasses the effect of training DS-TDNN with its HiddenSpeaker-protected dataset. This suggests that SLEM exhibits strong transferability, and the minDCF values for all three models exceeded 0.9, further supporting this observation.

In conclusion, SLEM demonstrates robust adaptability and portability, expanding its applicability to various scenarios. This undoubtedly strengthens the possibility of utilizing SLEM in privacy protection within complex real-world scenarios.

\section*{Conclusion}
In this paper, we explore the possibility of preventing personal speech data from unauthorized exploitation for speaker verification model training by introducing noise that is difficult for humans to capture. We propose a system called HiddenSpeaker. In this system, we introduce a Single-Level Error-Minimizing noise and demonstrate its effectiveness across SOTA models. Additionally, we use a Perceptual Hybrid Losses function to make the noise imperceptible, enabling the practical application of SLEM in real-life scenarios. To ensure that the PHL does not impact the effectiveness of SLEM, we choose to add noise in the high values of amplitude that models are more inclined to capture. We validate the system's effectiveness and transferability through a series of experiments, representing a breakthrough in privacy protection for speaker verification. Our work still faces constraints. Our audio employs a fixed-position clipping approach. Additionally, data augmentation techniques are often used to enhance speaker verification performance, so solutions to the complexities encountered in real-world scenarios will be considered in the future.

\bibliographystyle{IEEEtran}
\bibliography{IEEEabrv,REFERENCES}

\begin{thebibliography}{10}
\providecommand{\url}[1]{#1}
\csname url@samestyle\endcsname
\providecommand{\newblock}{\relax}
\providecommand{\bibinfo}[2]{#2}
\providecommand{\BIBentrySTDinterwordspacing}{\spaceskip=0pt\relax}
\providecommand{\BIBentryALTinterwordstretchfactor}{4}
\providecommand{\BIBentryALTinterwordspacing}{\spaceskip=\fontdimen2\font plus
\BIBentryALTinterwordstretchfactor\fontdimen3\font minus \fontdimen4\font\relax}
\providecommand{\BIBforeignlanguage}[2]{{%
\expandafter\ifx\csname l@#1\endcsname\relax
\typeout{** WARNING: IEEEtran.bst: No hyphenation pattern has been}%
\typeout{** loaded for the language `#1'. Using the pattern for}%
\typeout{** the default language instead.}%
\else
\language=\csname l@#1\endcsname
\fi
#2}}
\providecommand{\BIBdecl}{\relax}
\BIBdecl

\bibitem{desplanques20_interspeech}
B.~Desplanques, J.~Thienpondt, and K.~Demuynck, ``{ECAPA-TDNN: Emphasized Channel Attention, Propagation and Aggregation in TDNN Based Speaker Verification},'' in \emph{Proc. Interspeech 2020}, 2020, pp. 3830--3834.

\bibitem{9747384}
H.~Shen, Y.~Yang, G.~Sun, R.~Langman, E.~Han, J.~Droppo, and A.~Stolcke, ``Improving fairness in speaker verification via group-adapted fusion network,'' in \emph{ICASSP 2022 - 2022 IEEE International Conference on Acoustics, Speech and Signal Processing (ICASSP)}, 2022, pp. 7077--7081.

\bibitem{10289030}
S.~Saini and N.~Saxena, ``Speaker anonymity and voice conversion vulnerability: A speaker recognition analysis,'' in \emph{2023 IEEE Conference on Communications and Network Security (CNS)}, 2023, pp. 1--9.

\bibitem{panayotov2015librispeech}
V.~Panayotov, G.~Chen, D.~Povey, and S.~Khudanpur, ``Librispeech: an asr corpus based on public domain audio books,'' in \emph{2015 IEEE international conference on acoustics, speech and signal processing (ICASSP)}.\hskip 1em plus 0.5em minus 0.4em\relax IEEE, 2015, pp. 5206--5210.

\bibitem{veaux2017english}
C.~Veaux, J.~Yamagishi, K.~MacDonald \emph{et~al.}, ``English multi-speaker corpus for cstr voice cloning toolkit,'' \emph{ac. uk/jyamagis/page3/page58/page58. html,[Jan. 9, 2017]}, 2017.

\bibitem{nagrani2020voxceleb}
A.~Nagrani, J.~S. Chung, W.~Xie, and A.~Zisserman, ``Voxceleb: Large-scale speaker verification in the wild,'' \emph{Computer Speech \& Language}, vol.~60, p. 101027, 2020.

\bibitem{chung18b_interspeech}
J.~S. Chung, A.~Nagrani, and A.~Zisserman, ``{VoxCeleb2: Deep Speaker Recognition},'' in \emph{Proc. Interspeech 2018}, 2018, pp. 1086--1090.

\bibitem{tarun2023fast}
A.~K. Tarun, V.~S. Chundawat, M.~Mandal, and M.~Kankanhalli, ``Fast yet effective machine unlearning,'' \emph{IEEE Transactions on Neural Networks and Learning Systems}, 2023.

\bibitem{ge2022wavefuzz}
Y.~Ge, Q.~Wang, J.~Zhang, J.~Zhou, Y.~Zhang, and C.~Shen, ``Wavefuzz: A clean-label poisoning attack to protect your voice,'' \emph{arXiv preprint arXiv:2203.13497}, 2022.

\bibitem{huang2021unlearnable}
H.~Huang, X.~Ma, S.~M. Erfani, J.~Bailey, and Y.~Wang, ``Unlearnable examples: Making personal data unexploitable,'' in \emph{International Conference on Learning Representations}, 2021.

\bibitem{benesty2011speech}
J.~Benesty, J.~Chen, and E.~A. Habets, \emph{Speech enhancement in the STFT domain}.\hskip 1em plus 0.5em minus 0.4em\relax Springer Science \& Business Media, 2011.

\bibitem{jensen2016algorithm}
J.~Jensen and C.~H. Taal, ``An algorithm for predicting the intelligibility of speech masked by modulated noise maskers,'' \emph{IEEE/ACM Transactions on Audio, Speech, and Language Processing}, vol.~24, no.~11, pp. 2009--2022, 2016.

\bibitem{liu2023inplace}
J.~Liu and X.~Zhang, ``Inplace cepstral speech enhancement system for the icassp 2023 clarity challenge,'' in \emph{ICASSP 2023-2023 IEEE International Conference on Acoustics, Speech and Signal Processing (ICASSP)}.\hskip 1em plus 0.5em minus 0.4em\relax IEEE, 2023, pp. 1--2.

\bibitem{wang2020high}
H.~Wang, X.~Wu, Z.~Huang, and E.~P. Xing, ``High-frequency component helps explain the generalization of convolutional neural networks,'' in \emph{Proceedings of the IEEE/CVF conference on computer vision and pattern recognition}, 2020, pp. 8684--8694.

\bibitem{abadi2016deep}
M.~Abadi, A.~Chu, I.~Goodfellow, H.~B. McMahan, I.~Mironov, K.~Talwar, and L.~Zhang, ``Deep learning with differential privacy,'' in \emph{Proceedings of the 2016 ACM SIGSAC conference on computer and communications security}, 2016, pp. 308--318.

\bibitem{gilad2016cryptonets}
R.~Gilad-Bachrach, N.~Dowlin, K.~Laine, K.~Lauter, M.~Naehrig, and J.~Wernsing, ``Cryptonets: Applying neural networks to encrypted data with high throughput and accuracy,'' in \emph{International conference on machine learning}.\hskip 1em plus 0.5em minus 0.4em\relax PMLR, 2016, pp. 201--210.

\bibitem{lai2022efficient}
R.~Lai, X.~Fang, P.~Zheng, H.~Liu, W.~Lu, and W.~Luo, ``Efficient fragile privacy-preserving audio watermarking using homomorphic encryption,'' in \emph{International Conference on Artificial Intelligence and Security}.\hskip 1em plus 0.5em minus 0.4em\relax Springer, 2022, pp. 373--385.

\bibitem{zorarpaci2020hybrid}
E.~ZORARPACI and S.~A. {\"O}zel, ``A hybrid approach of homomorphic encryption and differential privacy for privacy preserving classification,'' \emph{International Journal of Applied Mathematics Electronics and Computers}, vol.~8, no.~4, pp. 138--147, 2020.

\bibitem{wang2021infoscrub}
H.-P. Wang, T.~Orekondy, and M.~Fritz, ``Infoscrub: Towards attribute privacy by targeted obfuscation,'' in \emph{Proceedings of the IEEE/CVF Conference on Computer Vision and Pattern Recognition}, 2021, pp. 3281--3289.

\bibitem{10095173}
M.~Tran and M.~Soleymani, ``A speech representation anonymization framework via selective noise perturbation,'' in \emph{ICASSP 2023 - 2023 IEEE International Conference on Acoustics, Speech and Signal Processing (ICASSP)}, 2023, pp. 1--5.

\bibitem{mivule2013utilizing}
K.~Mivule, ``Utilizing noise addition for data privacy, an overview,'' \emph{arXiv preprint arXiv:1309.3958}, 2013.

\bibitem{li2023make}
X.~Li and M.~Liu, ``Make text unlearnable: Exploiting effective patterns to protect personal data,'' in \emph{Proceedings of the 3rd Workshop on Trustworthy Natural Language Processing (TrustNLP 2023)}, 2023, pp. 249--259.

\bibitem{barreno2006can}
M.~Barreno, B.~Nelson, R.~Sears, A.~D. Joseph, and J.~D. Tygar, ``Can machine learning be secure?'' in \emph{Proceedings of the 2006 ACM Symposium on Information, computer and communications security}, 2006, pp. 16--25.

\bibitem{biggio2012poisoning}
B.~Biggio, B.~Nelson, and P.~Laskov, ``Poisoning attacks against support vector machines,'' in \emph{Proceedings of the 29th International Coference on International Conference on Machine Learning}, 2012, pp. 1467--1474.

\bibitem{carlini2017towards}
N.~Carlini and D.~Wagner, ``Towards evaluating the robustness of neural networks,'' in \emph{2017 ieee symposium on security and privacy (sp)}.\hskip 1em plus 0.5em minus 0.4em\relax Ieee, 2017, pp. 39--57.

\bibitem{goldblum2022dataset}
M.~Goldblum, D.~Tsipras, C.~Xie, X.~Chen, A.~Schwarzschild, D.~Song, A.~Madry, B.~Li, and T.~Goldstein, ``Dataset security for machine learning: Data poisoning, backdoor attacks, and defenses,'' \emph{IEEE Transactions on Pattern Analysis and Machine Intelligence}, vol.~45, no.~2, pp. 1563--1580, 2022.

\bibitem{shafahi2018poison}
A.~Shafahi, W.~R. Huang, M.~Najibi, O.~Suciu, C.~Studer, T.~Dumitras, and T.~Goldstein, ``Poison frogs! targeted clean-label poisoning attacks on neural networks,'' \emph{Advances in neural information processing systems}, vol.~31, 2018.

\bibitem{guo2023masterkey}
H.~Guo, X.~Chen, J.~Guo, L.~Xiao, and Q.~Yan, ``Masterkey: Practical backdoor attack against speaker verification systems,'' in \emph{Proceedings of the 29th Annual International Conference on Mobile Computing and Networking}, 2023, pp. 1--15.

\bibitem{10082866}
Y.~Ge, Q.~Wang, J.~Yu, C.~Shen, and Q.~Li, ``Data poisoning and backdoor attacks on audio intelligence systems,'' \emph{IEEE Communications Magazine}, vol.~61, no.~12, pp. 176--182, 2023.

\bibitem{aghakhani2023venomave}
H.~Aghakhani, L.~Sch{\"o}nherr, T.~Eisenhofer, D.~Kolossa, T.~Holz, C.~Kruegel, and G.~Vigna, ``Venomave: Targeted poisoning against speech recognition,'' in \emph{2023 IEEE Conference on Secure and Trustworthy Machine Learning (SaTML)}.\hskip 1em plus 0.5em minus 0.4em\relax IEEE, 2023, pp. 404--417.

\bibitem{reynolds2000speaker}
D.~A. Reynolds, T.~F. Quatieri, and R.~B. Dunn, ``Speaker verification using adapted gaussian mixture models,'' \emph{Digital signal processing}, vol.~10, no. 1-3, pp. 19--41, 2000.

\bibitem{alam2011multi}
M.~J. Alam, P.~Kenny, P.~Ouellet, D.~O'Shaughnessy \emph{et~al.}, ``Multi-taper mfcc features for speaker verification using i-vectors,'' in \emph{2011 IEEE Workshop on Automatic Speech Recognition \& Understanding}.\hskip 1em plus 0.5em minus 0.4em\relax IEEE, 2011, pp. 547--552.

\bibitem{matvejka2011full}
P.~Mat{\v{e}}jka, O.~Glembek, F.~Castaldo, M.~J. Alam, O.~Plchot, P.~Kenny, L.~Burget, and J.~{\v{C}}ernocky, ``Full-covariance ubm and heavy-tailed plda in i-vector speaker verification,'' in \emph{2011 IEEE international conference on acoustics, speech and signal processing (ICASSP)}.\hskip 1em plus 0.5em minus 0.4em\relax IEEE, 2011, pp. 4828--4831.

\bibitem{chen2022large}
Z.~Chen, S.~Chen, Y.~Wu, Y.~Qian, C.~Wang, S.~Liu, Y.~Qian, and M.~Zeng, ``Large-scale self-supervised speech representation learning for automatic speaker verification,'' in \emph{ICASSP 2022-2022 IEEE International Conference on Acoustics, Speech and Signal Processing (ICASSP)}.\hskip 1em plus 0.5em minus 0.4em\relax IEEE, 2022, pp. 6147--6151.

\bibitem{variani2014deep}
E.~Variani, X.~Lei, E.~McDermott, I.~L. Moreno, and J.~Gonzalez-Dominguez, ``Deep neural networks for small footprint text-dependent speaker verification,'' in \emph{2014 IEEE international conference on acoustics, speech and signal processing (ICASSP)}.\hskip 1em plus 0.5em minus 0.4em\relax IEEE, 2014, pp. 4052--4056.

\bibitem{bai2021speaker}
Z.~Bai and X.-L. Zhang, ``Speaker recognition based on deep learning: An overview,'' \emph{Neural Networks}, vol. 140, pp. 65--99, 2021.

\bibitem{deng2019arcface}
J.~Deng, J.~Guo, N.~Xue, and S.~Zafeiriou, ``Arcface: Additive angular margin loss for deep face recognition,'' in \emph{Proceedings of the IEEE/CVF conference on computer vision and pattern recognition}, 2019, pp. 4690--4699.

\bibitem{li2023dual}
Y.~Li and X.~Lin, ``Dual-stream time-delay neural network with dynamic global filter for speaker verification,'' \emph{arXiv preprint arXiv:2303.11020}, 2023.

\bibitem{zhang22h_interspeech}
Y.~Zhang, Z.~Lv, H.~Wu, S.~Zhang, P.~Hu, Z.~Wu, H.~yi~Lee, and H.~Meng, ``{MFA-Conformer: Multi-scale Feature Aggregation Conformer for Automatic Speaker Verification},'' in \emph{Proc. Interspeech 2022}, 2022, pp. 306--310.

\bibitem{liu2022mfa}
T.~Liu, R.~K. Das, K.~A. Lee, and H.~Li, ``Mfa: Tdnn with multi-scale frequency-channel attention for text-independent speaker verification with short utterances,'' in \emph{ICASSP 2022-2022 IEEE International Conference on Acoustics, Speech and Signal Processing (ICASSP)}.\hskip 1em plus 0.5em minus 0.4em\relax IEEE, 2022, pp. 7517--7521.

\end{thebibliography}

\end{document}